\documentclass[11pt]{article}               \def\new#1\endnew{{\bf #1}}  

\usepackage{latexsym,amsmath,amssymb,cite}

\let\bra=\langle        \let\ket=\rangle 

\newcommand {\ud} {\mathrm{d}}
\newcommand {\cF} {{\cal F}}

\newcommand {\cO}{{\cal O}}

\newcommand {\bbR}{\mathbb{R}}
\newcommand {\bbZ}{\mathbb{Z}}

\newcommand {\Back}{\!\!\!\!\!}

\newcommand {\Li} {\mathrm{Li}_2}

\newcommand {\tri}[2] {{}_{#1}\Delta_{#2}}
\newcommand {\tria} {~\Delta~}

\def\fnote#1#2{\begingroup\def\thefootnote{#1}\footnote{#2}
                \addtocounter{footnote}{-1}\endgroup}

\begin{document}                \baselineskip=16pt
\thispagestyle{empty}

\begin{flushright}
TUW--02--06\\
ITP--UH--04/02\\
hep-th/0203077
\end{flushright}

\vspace*{1.5cm}

\begin{center}
{\LARGE 
{\bf Non-commutative tachyon action and D-brane geometry
}}
\end{center}

\vspace*{5truemm}

\begin{center}
{\large Manfred Herbst,\fnote{\#}{e-mail: Manfred.Herbst@cern.ch}
        Alexander Kling,\fnote{$\,\Box$\,}{e-mail: kling@itp.uni-hannover.de}
        Maximilian Kreuzer\fnote{\,*\,}{e-mail: kreuzer@hep.itp.tuwien.ac.at}
}
\end{center}

{\sl 
\begin{center}
$^{\#}$ Theory Division, CERN\\
CH-1211, Geneva 23, Switzerland\\ 
{$^{\,\Box}$\,}Institut f\"ur Theoretische Physik, Universit\"at Hannover,\\
Appelstra\ss e 2, D-30167 Hannover, Germany\\
$^{*}$Institut f\"ur Theoretische Physik, Technische Universit\"at Wien,\\
Wiedner Hauptstra\ss e 8-10, A-1040 Vienna, Austria 

\end{center}
}

\vspace*{5mm}

\begin{abstract}        \normalsize
We analyse open 
string correlators in non-constant background fields, including the 
metric $g$, the antisymmetric $B$-field, and the gauge field $A$. 
Working with a derivative expansion for the background fields, but exact 
in their constant parts, we obtain a tachyonic on-shell condition for the
inserted functions and extract the kinetic term for the tachyon action. 
The 3-point correlator yields a non-commutative tachyon potential. We 
also find a remarkable feature of the differential structure on the
D-brane: Although the boundary metric $G$ plays an essential role in the
action, the natural connection on the D-brane is the same as in closed
string theory, i.e. it is compatible with the bulk metric and has
torsion $H=dB$. This means, in particular, that the parallel transport
on the brane is independent of the gauge field $A$.
\end{abstract}
\vfill

Keywords: Bosonic Strings, D-branes, Non-commutative Geometry \\[7pt]

\clearpage
\setcounter{page}{1}

\section{Introduction}
\label{sec:intro}

The geometry on a D-brane has recently attracted much attention, as it 
turned out to involve non-commutative structures that depend on 
the gauge invariant combination $\cF = B + (2\pi\alpha')dA$ of the bulk 
$B$-field and the boundary $A$-field.
A lot of work was done in examining the effect of a constant
$B$-field in a topological decoupling limit
\cite{Schomerus:1999ug,Seiberg:1999vs,Ardalan:1999ce,Ardalan:2000av,Chu:1999qz}.

A generalization to non-constant fields was given through the 
deformation quantization of Poisson manifolds. In that
case, a non-commutative product can be constructed to all orders
of derivatives out of the Poisson structure $\Theta$ and it 
represents the most general form being associative
\cite{Kontsevich:1997vb}. In open string 
theory this product appears in the decoupling limit when the $B$-field
(or equivalently $\cF$) is closed \cite{Cattaneo:2000fm}.

While the closure condition is necessary for associativity, it is
not required by
string theory and one may ask how far one can relax it in order to 
obtain a reasonable product. In \cite{Herbst:2001ai}
the non-commutative product was extracted from open string off-shell
correlators with insertions on the boundary of the disk.
It turned out that one has to abandon the decoupling limit in order to
retain a consistent setup. The only physical 
condition on the non-commutative parameter $\Theta$
in first derivative order of the background fields is the on-shell 
condition for the open
string gauge field $A$ on the D-brane, i.e. the generalized Maxwell
equation; see also \cite{Cornalba:2002sm,Ho:2000fv,Chu:2000gi,Ho:2001qk} 
for other attempts of treating a general background field $B$.

Imposing this equation has the
following consequences for the product in first derivative order 
\cite{Herbst:2001ai}.
Firstly, the non-commutative product of two functions equals the
ordinary product under the integral.\footnote{
In \cite{Herbst:2001ai} it was only shown that 
$\int f\circ g = \int g\circ f$, 
but it is easily checked that, in fact, $\int f\circ g = \int fg$.
}
Secondly, the product is associative up to a surface term.
As an immediate consequence, the product of 
an arbitrary number of functions is invariant under cyclic
permutations under the integral up to a possible change in the
bracket structure). The integration measure plays an important role 
in this respect and is given by a Born-Infeld measure, 
$\sqrt{\det(g-\cF)}$. Associativity cannot be maintained and must be 
replaced by an $A_\infty$-structure \cite{Cornalba:2002sm,stasheff}. 
Both properties are necessary to construct a reasonable action and a 
variation principle in terms of the non-commutative
product, the second one, to adjust the position of the variation of
the field and the first one, to remove all derivatives from the variation.

So far only the non-commutative product arising in this generalized setting
was considered. It can be
extracted purely from off-shell correlation functions. However, since 
string vacua correspond to 2-d 
conformal field theories, the correlators must finally take 
shape of the usual simple on-shell form \cite{Polchinski:1998rq}. 
There is some interesting information which one can gain from passing 
to on-shell correlators and it is the intention of this 
article to work out this information. 

For this purpose we will use several results from \cite{Herbst:2001ai}
and, therefore, inherit the
general setting of the model considered there. The open strings move
in a background including a general metric $g$ and a nontrivial $B$-field 
in the bulk and a gauge field $A$ on the boundary of the world sheet.
The world sheet is taken to be the upper half complex plane. All 
information is extracted from correlation functions using a derivative 
expansion of the background fields, where the expansion is restricted to 
first derivative order, but exact to all orders in the constant part of 
$\cF$.

\newpage
\noindent
What can we learn from the on-shell correlators? 

(i) As the insertions at the boundary of the disk are taken to 
be ordinary functions of the target space coordinates $X^\mu$ we
expect that the on-shell condition is tachyonic. 
Furthermore, the equation of motion is linear in the tachyonic field
since the insertions represent asymptotic states. We will deduce this
linear equation from the requirement that the on-shell correlators
must have the CFT form and thus obtain 
the kinetic term for the effective action of the tachyon. 

(ii) The explicit form of the on-shell three-point function then gives 
us the cubic interaction of the open string tachyon potential. 
Higher $n$-point correlators are difficult to manage. Nevertheless, 
in view of the cyclicity of the product under the integral, we are able to 
discuss some implications for the structure of higher order 
interactions.

(iii) Working in first derivative order of the background fields is
already sufficient to extract information about the differential structure 
from the tachyon equation of motion. As an interesting and somewhat 
surprising result we anticipate that, using the generalized Maxwell equation, 
the covariant derivative on the D-brane turns out to be the same as that off 
the brane, i.e. it is the connection compatible with the bulk metric and 
with torsion $H=dB$.

The organisation of the paper is as follows.
We start in section \ref{sec:NCP} with the introduction of some notation 
and review the properties of the non-commutative product. In
section \ref{sec:offshell} we calculate the full two- and three-point
off-shell correlators using results from
\cite{Herbst:2001ai}. Thereafter, in section \ref{sec:onshell}, we show
that conformal invariance requires a tachyonic on-shell condition for
the insertions of the correlators and the use of the Maxwell equation for 
the background fields. Eventually, we investigate the potential and the 
differential structure of the tachyonic action in section \ref{sec:tachyon} 
and close with a general discussion of our results in section \ref{sec:disc}.

\section{The non-commutative product}
\label{sec:NCP}

On the D-brane we have in addition to the so called bulk metric 
$g_{\mu\nu}$, which enters in the sigma model action, the
boundary metric $G^{\mu\nu}$. Two metrics arise because of the fact that
on the brane one has to consider the combination
$M_{\mu\nu} = g_{\mu\nu} + \cF_{\mu\nu}$ rather than the separate
quantities $g$ and $\cF$. Consequently, one can split the inverse of $M$
into the symmetric and antisymmetric part, i.e. 
$M^{\mu\nu} := M^{-1}{}^{\mu\nu} = G^{\mu\nu} + \Theta^{\mu\nu}$, and obtains
the second metric $G$ and the antisymmetric part $\Theta$, which turns out
to be the non-commutativity parameter.\footnote{%
  We use the convention $M^{-1}{}^{\mu\nu} M_{\nu\rho} = \delta^\mu{}_\rho$.
}

The product found in \cite{Herbst:2001ai} is given to all orders in
$\Theta$ and to
first derivative order in the background fields. It reads
\begin{eqnarray}
  \label{eq:NCP}
  f(x) \; \circ \; g(x) \;=\; f * g 
  \Back&-&\Back \frac 1{12}
  \Theta^{\mu\rho} \partial_\rho \Theta^{\nu\sigma} \;
  \Bigl(\partial_\mu \partial_\nu f \;* \partial_\sigma g +
  \partial_\sigma f \;* \partial_\mu \partial_\nu g \Bigr) +\nonumber\\
  &+& \cO\bigl((\partial\Theta)^2,\partial^2\Theta\bigr) ,
\end{eqnarray}
where '$*$' denotes the Moyal contribution to the product. The
on-shell condition for the gauge field $A$ on the D-brane is
\begin{eqnarray}
  \label{eq:maxwell}
  G^{\rho\sigma}D_\rho \cF_{\sigma\mu}
  -\frac 12 \Theta^{\rho\sigma}H_{\rho\sigma}{}^\lambda \cF_{\lambda\mu} = 0,
\end{eqnarray}
where $D_\rho$ is the Christoffel connection of $g$, or equivalently,
\begin{eqnarray}
  \label{eq:eom1}
  \partial_\mu \Bigl(\sqrt{g-\cF}~\Theta^{\mu\nu}\Bigr) &=& 0.
\end{eqnarray}
Imposing this equation of motion one finds 
that the product of
two functions equals the ordinary product under the integral,
\begin{eqnarray}
  \label{eq:ordprod}
  \int \ud^{D} x \sqrt {\det(g\!\!-\!\!\cF)} ~f \circ g &=& 
  \int \ud^{D} x \sqrt {\det(g\!\!-\!\!\cF)} ~f \cdot g ~,
\end{eqnarray}
and 
that it is associative up to a surface term,
\begin{eqnarray}
  \label{eq:assocprod}
  \int \ud^{D} x \sqrt {\det(g\!\!-\!\!\cF)} ~(f \circ  g) \circ h &=& 
  \int \ud^{D} x \sqrt {\det(g\!\!-\!\!\cF)} ~ f \circ (g \circ  h) ~.
\end{eqnarray}
The trace property
\begin{eqnarray}
  \label{eq:trace}
  \int \ud^{D} x \sqrt {\det(g\!\!-\!\!\cF)} 
  \Bigl((...( f_1 \circ ...))\circ f_{n\!-\!1}\Bigr) \circ f_n = 
  \int \ud^{D} x \sqrt {\det(g\!\!-\!\!\cF)} 
  \Bigl(f_n \circ (...( f_1 \circ ... ))\Bigr)\circ f_{n\!-\!1} ~.
\end{eqnarray}
follows immediately. Although the integration measure plays an important
role in order to derive
these properties we will subsequently use the abbreviation
$\int _x = \int \ud^{D} x \sqrt {\det(g\!\!-\!\!\cF)}$ for the integral.
These results will extensively be used in sections
(\ref{sec:onshell}) and (\ref{sec:tachyon}).

\section{Off-shell correlators}
\label{sec:offshell}

We are now going to use several results of the appendix of 
\cite{Herbst:2001ai} to calculate the full off-shell two- and three-point
correlators of ordinary functions of target space coordinates $X^\mu$. 
The insertions are ordered at the boundary of the upper half plane, so
that $\tau_1 < \tau_2 < \tau_3$.

\paragraph{The two-point correlator} \hspace*{-0.3cm}was already given in
\cite{Herbst:2001ai} and we repeat the expression in a more compact form. 
To this end we use the relation
\begin{eqnarray}
  \label{eq:Gconnect}
  2 G^{\nu\rho} G^{\sigma\lambda} \bar \Gamma_{\rho\lambda}{}^{\mu} =
  G^{\mu\rho} \partial_\rho G^{\nu\sigma}-
  G^{\nu\rho} \partial_\rho G^{\sigma\mu}-
  G^{\sigma\rho} \partial_\rho G^{\nu\mu}~,
\end{eqnarray}
to introduce the Christoffel connection compatible with $G$. 
With the abbreviation ($\tau_{ij} = \tau_i - \tau_j$)
\begin{eqnarray}
  \label{eq:triangle}
  f \tria g := \sum _{n=0}^\infty 
             \frac 1{n!} \Bigl(\frac{1}{2\pi}\Bigr)^n
             \ln^n\tau_{21}^{-2}~G^{\mu^n\nu^n} \;
             \bar D_{\mu^n} f \; \circ \;
             \bar D_{\nu^n}g
\end{eqnarray}
where the upper subindices of the indices mean a product of derivatives,
$\bar D_{\mu^n} := \bar D_{(\mu_1} \ldots \bar D_{\mu_n)} =\partial_{\mu^n} f - \frac {n(n-1)}{2}~\bar\Gamma_{(\mu_1\mu_2}{}^\rho \partial_{\mu_3}\ldots\partial_{\mu_n)} \partial_{\rho} f + \cO(\partial^2)$
and a product of metrics
$G^{\mu^n\nu^n}=G^{\mu_1\nu_1}\ldots G^{\mu_n\nu_n}$. The symmetrization
of the derivatives $\bar D_{\mu}$ in $\bar D_{\mu^n}$
is automatic in first order, because the partial derivatives contracted with
$G$ symmetrize the $\bar\Gamma$-term on the other side, i.e.
$G^{\mu\nu} \bar D_\mu f \bar D_\nu g = G^{\mu\nu} \partial_\mu f \bar D_\nu g + G^{\mu\nu} \bar D_\mu f \partial_\nu g + \cO(\partial^2)$. 
We can then write the full two-point correlator as
\begin{eqnarray}
  \label{eq:2pCorr}
  \bra\;:\!f[X(\tau_1)]\!: \, :\!g[X(\tau_2)]\!:\;\ket
      \; &=& 
  \int _x f \tria g \nonumber \\
  + \frac i{4\pi} \ln \tau_{21}^{-1}\int _x 
              \Theta^{\mu\rho}\partial_\rho G^{\nu\sigma} 
              \Back &\bigl(& \Back
                    \partial_\nu \partial_\sigma f \tria \partial_\mu g -
                    \partial_\mu f \tria \partial_\nu \partial_\sigma g
              \bigr)  
\nonumber \\
  + \frac i{2\pi} \ln \tau_{21}^{-1}\int _x 
              G^{\nu\sigma}\partial_\sigma \Theta^{\rho\mu}
              \Back &\bigl(& \Back
                    \partial_\nu \partial_\rho f \tria \partial_\mu g -
                    \partial_\mu f \tria \partial_\nu \partial_\rho g
              \bigr)
  + \cO(\partial^2).
\end{eqnarray}

\paragraph{The three-point correlator}\hspace*{-0.3cm}
is much more complicated. It is a
rather tedious but straightforward work to collect all the 
relevant terms from the appendix in \cite{Herbst:2001ai}. 
In order to realize the structure more clearly we
first consider the two cases, $\Theta = 0$ and $\Theta \rightarrow \infty$
(or, equivalently, $G \rightarrow 0$). 
For $\Theta = 0$, the covariant derivative $\bar D$ again appears, now
in the combination
\begin{eqnarray}
  \label{eq:DoubleDer}
  \bar D_{\mu^n} \bar D_{\nu^m} f &=& 
                     \partial_{\mu^n}\partial_{\nu^m} f
\nonumber\\
                 &-& \frac {n(n-1)}{2}~\bar\Gamma_{(\mu_1\mu_2}{}^\rho
                     \partial_{\mu_3}\ldots\partial_{\mu_n)}
                     \partial_{\nu^m} \partial_{\rho} f +
\nonumber\\
                 &-& \frac {m(m-1)}{2}~\bar\Gamma_{(\nu_1\nu_2}{}^\rho
                     \partial_{\nu_3}\ldots\partial_{\nu_m)}
                     \partial_{\mu^n} \partial_{\rho} f +
\nonumber\\
                 &-& mn~\bar\Gamma_{(\mu_1|(\nu_1}{}^\rho
                     \partial_{\nu_2}\ldots\partial_{\nu_m)}
                     \partial_{|\mu_2}\ldots\partial_{\mu_n)}
                     \partial_{\rho} f +
                     \cO(\partial^2).
\end{eqnarray}
The correlator is given as
\begin{eqnarray}
  \label{eq:3pCorrG}
  &&\bra\;:\!f[X(\tau_1)]\!: \, :\!g[X(\tau_2)]\!:\; :\!h[X(\tau_3)]\!:\;\ket
  \bigr|_{\Theta = 0} \; = 
  \int _x \tri{}{}(f,g,h) + \cO(\partial^2) =
\nonumber\\
  &&\sum _{I,J,K} \bigl( \frac 1\pi \bigr)^{I+J+K} 
  \frac {\ln^I\tau_{21}^{-1} \ln^J\tau_{31}^{-1} \ln^K\tau_{32}^{-1}}
        {I! J! K!} \times 
\nonumber\\
  &&\times\int _x G^{\mu^I\nu^I}G^{\mu^J\nu^J}G^{\mu^K\nu^K} 
         \bar D_{\mu^I\mu^J} f ~ \bar D_{\nu^I\mu^K} g ~ \bar D_{\nu^J\nu^K} h
  + \cO(\partial^2) .
\end{eqnarray}
The triangle functional $\tri{}{}(f,g,h)$ in the first line is a
generalization of (\ref{eq:triangle}) and was introduced for later
reference. 
In case of $G \rightarrow 0$ the correlator can be expressed 
solely through
the non-commutative product (\ref{eq:NCP}) (cf. also \cite{Cornalba:2002sm})
\begin{eqnarray}
  \label{eq:3pCorrTheta}
  \bra\;:\!f[X_1]\!: \, :\!g[X_2]\!:\; :\!h[X_3]\!:\;\ket
  \bigr|_{G = 0} &=& 
  \int _x \frac 12 \Bigl\{ (f \circ g) \circ h + f \circ (g \circ h)
\\
  &+& L(m)~\Bigl(f_1\circ (f_2\circ f_3) - (f_1\circ f_2)\circ f_3) 
  \Bigr) \Bigr\}
\nonumber\\
  &+& \cO(\partial^2).
\nonumber
\end{eqnarray}
$L(m) = \frac{6}{\pi^2}(\Li(m) - \Li(1-m))$ is an antisymmetric combination
of dilogarithms $\Li(m)$ with the limits $L(0) = -1$ and $L(1) = 1$.
The modulus $m = \tau_{21}/\tau_{31}$ can take the values $0 \leq m \leq 1$.

Since we take into account only terms to first derivative 
order these two results can easily be completed 
to the general case. In equations (\ref{eq:3pCorrG}) and
(\ref{eq:3pCorrTheta}) we have included the $G\partial G$- and the
$\Theta\partial\Theta$-terms, respectively. In the full correlator these
two results combine in a natural way and add up 
with the remaining $G\partial \Theta$ and $\Theta\partial G$ parts, so
that we find
\begin{eqnarray}
  \label{eq:3pCorr}
  \bra\;:\!f[X(\tau_1)]\!: \, :\!g[X(\tau_2)]\!:\; :\!h[X(\tau_3)]\!:\;\ket
  \; = \; F[G\partial G, \Theta\partial\Theta] + 
          F[G\partial\Theta, \Theta\partial G] + \cO(\partial^2)
\end{eqnarray}
where
\begin{eqnarray}
  \label{eq:GdGTdT}
  &&F[G\partial G, \Theta\partial\Theta] \; = \;\sum _{I,J,K} 
  \frac {\ln^I\tau_{21}^{-1} \ln^J\tau_{31}^{-1} \ln^K\tau_{32}^{-1}}
        {I! J! K!}\times
\\
  &\times& \frac 1{\pi^{I+J+K}} 
    \int _x G^{\mu^I\nu^I}G^{\mu^J\nu^J}G^{\mu^K\nu^K}
      \frac 12 \Bigl[ \bar D_{\mu^I\mu^J}f \circ 
      (\bar D_{\nu^I\mu^K}g \circ \bar D_{\nu^J\nu^K}h) +
      (\bar D_{\mu^I\mu^J}f \circ \bar D_{\nu^I\mu^K}g) 
      \circ \bar D_{\nu^J\nu^K}h 
\nonumber\\
  && \hspace*{3.5truecm}+~L(m)~\Bigl(\bar D_{\mu^I\mu^J}f\circ 
                 (\bar D_{\nu^I\mu^K}g\circ \bar D_{\nu^J\nu^K}h) - 
               (\bar D_{\mu^I\mu^J}f\circ \bar D_{\nu^I\mu^K}g)\circ 
                \bar D_{\nu^J\nu^K}h) 
  \Bigr) \Bigr]
\nonumber
\end{eqnarray}
and, with $\partial_\mu f = f_\mu$, 
\begin{eqnarray}
  \label{eq:GdTTdG}
  && F[G\partial\Theta, \Theta\partial G] \; = \;
\nonumber\\
   &+& \frac i{2\pi} \ln\tau_{21}^{-1}\int _x 
              \Theta^{\rho\sigma} \partial_\sigma G^{\mu\nu}
              [-         \tri{}{} (f_\mu, g_\nu, h_\rho)
               -\frac 12 \tri{}{} (f_\rho, g_{\mu\nu}, h)
               +\frac 12 \tri{}{} (f_{\mu\nu}, g_\rho, h)
              ]
\nonumber\\
   &+& \frac i{2\pi} \ln\tau_{31}^{-1}\int _x 
              \Theta^{\rho\sigma} \partial_\sigma G^{\mu\nu}
              [-\tri{}{} (f_\rho, g_\mu, h_\nu)
               +\tri{}{} (f_\mu, g_\nu, h_\rho)
               -\frac 12 \tri{}{} (f_\rho, g, h_{\mu\nu})
               +\frac 12 \tri{}{} (f_{\mu\nu}, g, h_\rho)
              ]
\nonumber\\
   &+& \frac i{2\pi} \ln\tau_{32}^{-1}\int _x 
              \Theta^{\rho\sigma} \partial_\sigma G^{\mu\nu}
              [+\tri{}{} (f_\rho, g_\mu, h_\nu)
               -\frac 12 \tri{}{} (f, g_\rho, h_{\mu\nu})
               +\frac 12 \tri{}{} (f, g_{\mu\nu}, h_\rho)
              ]
\nonumber\\
   &+& \frac i{2\pi} \ln\tau_{21}^{-1}\int _x 
              G^{\rho\sigma} \partial_\sigma \Theta^{\mu\nu}
              [+\tri{}{} (f_\mu, g_{\nu\rho}, h)
               -\tri{}{} (f_{\nu\rho}, g_\mu, h)
              ]
\nonumber\\
   &+& \frac i{2\pi} \ln\tau_{31}^{-1}\int _x 
              G^{\rho\sigma} \partial_\sigma \Theta^{\mu\nu}
              [+\tri{}{} (f_\mu, g_\nu, h_\rho)
               +\tri{}{} (f_\nu, g_\rho, h_\mu)
               -\tri{}{} (f_\rho, g_\mu, h_\nu)
\nonumber\\
   & & ~~~~~~~~~~~~~~~~~~~~~~~~~~~~~~ +\tri{}{} (f_\mu, g, h_{\nu\rho})
               -\tri{}{} (f_{\nu\rho}, g, h_\mu)
              ]
\nonumber\\
   &+& \frac i{2\pi} \ln\tau_{32}^{-1}\int _x 
              G^{\rho\sigma} \partial_\sigma \Theta^{\mu\nu}
              [+\tri{}{} (f, g_\mu, h_{\nu\rho})
               -\tri{}{} (f, g_{\nu\rho}, h_\mu)
              ].
\end{eqnarray}
The symbol $\tri{}{} (f, g, h)$ is defined in equation (\ref{eq:3pCorrG}).

\section{On-shell correlators}
\label{sec:onshell}

In the previous section the insertions on the disk as well as the 
background fields are completely general, they do not satisfy
any on-shell conditions, which are determined by the conformal
invariance of the theory. The equations of motion for the 
background fields are given by
the $\beta$-functions of the world sheet theory whereas
the equations of motion for the insertions are determined by the 
conformal transformation properties of the correlation functions.

The correlators of the CFT on the disk must be invariant under 
the global conformal group $SL(2,\bbR)$. In particular, the
2-point correlator with insertions
at the boundary is
\begin{eqnarray}
  \label{eq:CFT2point}
  <f_1[X(\tau_1)] f_2[X(\tau_2)]> =  \frac {C_{12}}{(\tau_{21})^{2h}},
\end{eqnarray}
where $h$ is the conformal weight of both $f_1$ and $f_2$.
The correlator for operators with different weights vanishes.
The 3-point correlator is
\begin{eqnarray}
  \label{eq:CFT3point}
   <f_1[X(\tau_1)] f_2[X(\tau_2)] f_3[X(\tau_3)]> =
   \frac {C_{123}}
   {\tau_{21}^{h_1+h_2-h_3}\tau_{31}^{h_3+h_1-h_2}\tau_{32}^{h_2+h_3-h_1}}
\end{eqnarray}
The constants $C_{12}$ and $C_{123}$ are functionals of $f_i(x)$,
independent of the positions $\tau_i$ and invariant under cyclic permutation
of indices. For physical fields which should carry $h = 1$ we have
\begin{eqnarray}
  \label{eq:CFT2corrh1}
  <f_1[X_1]~f_2[X_2]> &=&  \frac {C_{12}}{(\tau_1 - \tau_2)^{2}},
\\
  \label{eq:CFT3corrh1}
  <f_1[X_1]~f_2[X_2]~f_3[X_3]> &=&
  \frac {C_{123}}{\tau_{21}\tau_{31}\tau_{32}}.
\end{eqnarray}

On the other hand, the off-shell correlators
in the open string background, (\ref{eq:2pCorr}) and (\ref{eq:3pCorr}),
can be written in the following way,
\begin{eqnarray}
  \label{eq:2pointoff}
  <f_1[X_1]~f_2[X_2]> &=&
  \sum _{I=0}^{\infty} \frac {\ln^I\tau_{21}^{-2}}{I!}  F_I[f_i](\tau_i)
\\
  <f_1[X_1]~f_2[X_2]~f_3[X_3]> &=&
  \sum _{I,J,K=0}^{\infty}
  \frac{\ln^I\tau_{21}^{-1}\ln^J\tau_{31}^{-1}\ln^K\tau_{32}^{-1}}{I!J!K!} 
  F_{IJK}[f_i](\tau_i)
\nonumber
\end{eqnarray}
where $F_I[f_i](\tau_i)$ and $F_{IJK}[f_i](\tau_i)$ are functionals of
$f_i$ and functions of $\tau_i$. The $\tau_i$-dependence arises from
the dilogarithm in (\ref{eq:GdGTdT}) as well as from the sign
function $\epsilon(\tau_{ij})$ which
accompanies every $\Theta$ (cf. \cite{Herbst:2001ai}). 
In fact, we do not see the sign function
because of our choice of ordering, $\tau_1 < \tau_2 < \tau_3$.

Therefore, if 
\begin{eqnarray}
  \label{eq:reduction1}
  F_I[f_i](\tau_i) &=& F_{I-1}[f_i](\tau_i) 
\\
  \label{eq:reduction2}
  F_{IJK}[f_i](\tau_i) &=& F_{(I-1)JK}[f_i](\tau_i) = 
  F_{I(J-1)K}[f_i](\tau_i) = F_{IJ(K-1)}[f_i](\tau_i)
\end{eqnarray}
is fulfilled, one can reduce all functionals in the sum to
$F_0[f_i](\tau_i)$ and $F_{000}[f_i](\tau_i)$, respectively.
Furthermore, in order to reproduce the behaviour (\ref{eq:CFT2corrh1})
and (\ref{eq:CFT3corrh1}), $F_0[f_i](\tau_i)$ and $F_{000}[f_i](\tau_i)$
must be constants and then determine $C_{12}=F_0[f_i]$ and
$C_{123}=F_{000}[f_i]$.
However, this does not work off-shell and has to be accomplished by
certain on-shell conditions imposed  on the insertions 
(and of course on the background fields). We proceed in two steps and 
first show the following theorem:
\paragraph{Relations}\hspace*{-0.3cm}
(\ref{eq:reduction1}) and (\ref{eq:reduction2}) require 
that the insertions satisfy the tachyonic equation of
motion
\begin{eqnarray}
  \label{eq:eomtachyon}
  \square f_i - (-2\pi) f_i = 
  \frac {1}{\sqrt{\det(g-\cF)}} ~\partial_\mu 
  \bigl(\sqrt{\det(g-\cF)}G^{\mu\nu}\partial_\nu f_i \bigr) - (-2\pi) f_i = 0.
\end{eqnarray}

\noindent
\emph{Proof:} We start with the 2-point correlator. In order to get a
scalar equation of the insertions one has to integrate by part. 
On the other hand, if we look at S-matrix calculations 
\cite{Polchinski:1998rq} the momentum 
conservation ($\delta^D(\Sigma k)$) comes from the integration over
zero modes. Here we do not have a flat background and we cannot perform
the integration in that way. Now the integration by part is the analog of
the momentum conservation in position space. Furthermore, one can
separate the functionals $F_I$ into two distinct parts,
$F_I = F_I[\Theta d\Theta] + F_I[GdG,Gd\Theta,\Theta dG]$. The former
term comes from the first line of (\ref{eq:2pCorr}), but without the
Christoffel symbols, the latter arises from the rest of (\ref{eq:2pCorr}).

We take $F_I[\Theta d\Theta]$ and integrate by part in the following way
\begin{eqnarray}
  \label{eq:intpart}
  F_I[\Theta d\Theta] &=& \Bigl(\frac{1}{2\pi}\Bigr)^I \int _x G^{\mu^I\nu^I}
          \partial_{\mu^I} f_1 \; \circ \;\partial_{\nu^I}f_2
\\
    =  &-& \frac 12 \Bigl(\frac{1}{2\pi}\Bigr)^I \int _x G^{\mu^{I-1}\nu^{I-1}}
          \partial_{\mu^{I-1}} 
          \square f_1
          \; \circ \; \partial_{\nu^{I-1}}f_2
\nonumber\\
      &-& \frac 12 \Bigl(\frac{1}{2\pi}\Bigr)^I \int _x G^{\mu^{I-1}\nu^{I-1}}
          \partial_{\mu^{I-1}} f_1
          \; \circ \; \partial_{\nu^{I-1}}
          \square f_2
\nonumber\\
      &+& F'_I[GdG,Gd\Theta,\Theta dG].
\nonumber
\end{eqnarray}
The last expression $F'_I[GdG,Gd\Theta,\Theta dG]$ combines with
$F_I[GdG,Gd\Theta,\Theta dG]$.
Analogously, using integration by part in 
$F''_I=F'_I[GdG,Gd\Theta,\Theta dG] + F_I[GdG,Gd\Theta,\Theta dG]$, 
the differential operator as it appears in (\ref{eq:intpart}) arises
now in zero derivative order of the background fields, i.e.
$\square f_i = G^{\mu_I\nu_I} \partial_{\mu_I}\partial_{\nu_I} f_i$,
because $F''_I$ contains only terms of first derivative
order. Now $F_{I-1}[\Theta d\Theta]$ is equal to the first two lines of
(\ref{eq:intpart}) if equation (\ref{eq:eomtachyon}) holds. Proceeding 
along the same lines one can show that 
$F_{I-1}[GdG,Gd\Theta,\Theta dG] = F''_I$,
again using (\ref{eq:eomtachyon}).
In fact, the 2-point function does not fix the tachyonic equation uniquely.
One can add $A^{\mu^{2n+1}} \partial_{\mu^{2n+1}}f_i$, where
$n \in \bbZ$ and $A \sim \cO(\partial)$. By means of partial
integration such terms would mutually cancel
in the second and third line of (\ref{eq:intpart}). However, we will see 
that this ambiguity is fixed by the 3-point correlator.

The calculation for the 3-point correlator is similar. First we make the
split $F_{IJK}=F_{IJK}[\Theta d\Theta] + F_{IJK}[GdG,Gd\Theta,\Theta dG]$.
Let us again look at the S-matrix calculation. There one uses the relation
$k_1 k_2 = \frac 14 (k_1-k_2-k_3)(k_2-k_1-k_3)=\frac 12(k_3^2 -k_1^2 -k_2^2)$.
With the analogous transformation in terms of partial integrations we obtain
from (\ref{eq:3pCorr}) and (\ref{eq:GdGTdT})
\begin{eqnarray}
  \label{eq:intpart3}
  &&F_{IJK}[\Theta d\Theta] =
\\ 
  &=&\frac 1{\pi^{I+J+K}} 
    \int _x G^{\mu^I\nu^I}G^{\mu^J\nu^J}G^{\mu^K\nu^K}
      \frac 12 \Bigl[ \partial_{\mu^I\mu^J}f_1 \circ 
      (\partial_{\nu^I\mu^K}f_2 \circ \partial_{\nu^J\nu^K}f_3) +
      (\textrm{other bracket})
\nonumber\\
  && \hspace*{3.5truecm}+~L(m)~\Bigl(\partial_{\mu^I\mu^J}f_1\circ 
                 (\partial_{\nu^I\mu^K}f_2\circ \partial_{\nu^J\nu^K}f_3) -
                 (\textrm{other bracket})
  \Bigr) \Bigr] =
\nonumber\\
  &=&\frac 1{\pi^{I+J+K}} 
    \int _x G^{\mu^{I-1}\nu^{I-1}}G^{\mu^J\nu^J}G^{\mu^K\nu^K}
      \frac 14 \Bigl[ 
      \partial_{\mu^{I-1}\mu^J}f_1 \circ 
      (\partial_{\nu^{I-1}\mu^K}f_2 \circ \partial_{\nu^J\nu^K}\square f_3)
\nonumber\\
  &&  \hspace*{6truecm}
      -\partial_{\mu^{I-1}\mu^J}\square f_1 \circ 
      (\partial_{\nu^{I-1}\mu^K}f_2 \circ \partial_{\nu^J\nu^K}f_3)
\nonumber\\
  &&  \hspace*{6truecm}
      -\partial_{\mu^{I-1}\mu^J}f_1 \circ 
      (\partial_{\nu^{I-1}\mu^K}\square f_2 \circ \partial_{\nu^J\nu^K}f_3)
\nonumber\\      
  &&  \hspace*{6truecm}
      +(\textrm{other bracket})
\nonumber\\
  && \hspace*{3.5truecm}+~L(m)~\Bigl(
      \partial_{\mu^{I-1}\mu^J}f_1\circ 
      (\partial_{\nu^{I-1}\mu^K}f_2\circ \partial_{\nu^J\nu^K}\square f_3)
\nonumber\\
  &&  \hspace*{5truecm}
      -\partial_{\mu^{I-1}\mu^J}\square f_1 \circ 
      (\partial_{\nu^{I-1}\mu^K}f_2 \circ \partial_{\nu^J\nu^K}f_3)
\nonumber\\
  &&  \hspace*{5truecm}
      -\partial_{\mu^{I-1}\mu^J}f_1 \circ 
      (\partial_{\nu^{I-1}\mu^K}\square f_2 \circ \partial_{\nu^J\nu^K}f_3)
\nonumber\\      
  &&  \hspace*{5truecm}
      -(\textrm{other bracket})
  \Bigr) \Bigr]
\nonumber\\
  && \hspace*{0.5truecm} +~F'_{IJK}[GdG,Gd\Theta,\Theta dG].
\nonumber
\end{eqnarray}
Since the dilogarithmic term is of first derivative order in background
fields there are no contributions
thereof in $F'_{IJK}[GdG,Gd\Theta,\Theta dG]$.
The same procedure as above shows that in view of (\ref{eq:eomtachyon})
the first part of (\ref{eq:intpart3}) equals $F_{(I-1)JK}[\Theta d\Theta]$
and $F'_{IJK}[GdG,Gd\Theta,\Theta dG] + F_{IJK}[GdG,Gd\Theta,
\Theta dG]= F_{(I-1)JK}[GdG,Gd\Theta,\Theta dG]$. But now terms like
$A^{\mu^{2n+1}} \partial_{\mu^{2n+1}}f_i$ would not cancel in
(\ref{eq:intpart3}), so that the tachyonic equation of motion
(\ref{eq:eomtachyon}) is unique. Relations $F_{IJK} = F_{I(J-1)K}$
and $F_{IJK} = F_{IJ(K-1)}$ can be shown analogously. $\qquad\square$

With this result we can write the correlators as
\begin{eqnarray}
  \label{eq:corrOnShell}
  <f_1[X_1]~f_2[X_2]> &=& \frac {F_0[f_i](\tau_i)}{\tau_{21}{}^2}
\\
  <f_1[X_1]~f_2[X_2]~f_3[X_3]> &=&
  \frac{F_{000}[f_i](\tau_i)}{\tau_{21}\tau_{31}\tau_{32}}
\nonumber  
\end{eqnarray}
with
\begin{eqnarray}
  \label{eq:corr2off}
  F_0[f_i](\tau_i) &=& \int _x (f_1\circ f_2)~,
\\
  \label{eq:corr3off}
   F_{000}[f_i](\tau_i) &=& \int _x \frac 12 \Bigl\{
          f_1\circ (f_2\circ f_3) + (f_1\circ f_2)\circ f_3 +
\\
      && +~L(m)~\Bigl(f_1\circ (f_2\circ f_3) - (f_1\circ f_2)\circ f_3)\Bigr)
\Bigr\}.
\nonumber
\end{eqnarray}

Indeed, (\ref{eq:corr2off}) and (\ref{eq:corr3off}) are not yet position
independent and invariant under cyclic exchange of the functions $f_i$.
Putting also the background fields on shell, i.e. using 
the Maxwell equation (\ref{eq:eom1}), we can take advantage of
the relations (\ref{eq:ordprod}) and (\ref{eq:assocprod}). So, we reach 
the final result
\begin{eqnarray}
  \label{eq:corr2on}
  <f_1[X_1]~f_2[X_2]> &=& \frac {1}{\tau_{21}{}^2} \int _x f_1\circ f_2 
           ~=~ \frac {1}{\tau_{21}{}^2} \int _x f_1\cdot f_2~,
\\
  \label{eq:corr3on}
  <f_1[X_1]~f_2[X_2]~f_3[X_3]> &=&  \frac {1}{\tau_{21}\tau_{31}\tau_{32}}
                     \int _x f_1\circ f_2\circ f_3~.
\nonumber
\end{eqnarray}

We close this section with a remark on the ghost fields, which we have 
totally excluded from our discussion so far.
On the disc we have three conformal killing vectors (forming the M\"obius
group $SL(2,\bbR)$) and therefore three of
the vertices in a correlator can be fixed in position and must be 
accompanied by a ghost field $c(\tau_i)$, the 
others being integrated over the world sheet. The 2-point correlator has too
few insertions in order to give a non-vanishing result in the ghost sector,
i.e. $\bra c(\tau_1)c(\tau_2) \ket_{gh} = 0$. The 3-point ghost amplitude, 
$\bra c(\tau_1)c(\tau_2)c(\tau_3) \ket_{gh} 
        = c_{gh}~\tau_{21}\tau_{31}\tau_{32}$,
exactly cancels the position dependence of the correlator 
(\ref{eq:CFT3corrh1}). Moreover, the M\"obius group preserves
the cyclic order of the insertions and so we must sum over inequivalent
orderings in the 3-point amplitude, so that we obtain\footnote{%
  Until now we have chosen the normalization of the correlators such
  that it reduces to an integral over the ordinary product of
  functions for
  $\cF=0$. We reintroduce a normalization constant $c = c_X c_{gh}$
  where $c_X$ and $c_{gh}$ are the normalizations for the matter
  and the ghost contribution, respectively. $c$ is fixed by unitarity
  \cite{Polchinski:1998rq}.
}
\begin{eqnarray}
  \label{eq:fullcorr}
  \bra cf_1[X_1]~cf_2[X_2]~cf_3[X_3]\ket + (f_2 \leftrightarrow f_3) =
  c~\int _x f_1\circ (f_2\circ f_3+f_3\circ f_2)~.
\end{eqnarray}

\section{Tachyonic action}
\label{sec:tachyon}

The results (\ref{eq:eomtachyon}) and (\ref{eq:fullcorr}) enable us to
reconstruct the kinetic term and the cubic potential of the open
string tachyon. The value for the coupling constant is recovered from 
consistency with S-matrix calculations, as discussed 
e.g. in \cite{Polchinski:1998rq}, by  
taking the limit $\Theta \rightarrow 0$ and $g_{\mu\nu} = \eta_{\mu\nu}$.
\begin{eqnarray}
  \label{eq:tachaction}
  S &=& - \frac 1{2g_o^2} \int d^Dx \sqrt{g-\cF}~
      \Bigl\{ 
      G^{\mu\nu} \cdot \partial_\mu T \cdot \partial_\nu T -
      \frac {1}{\alpha'} ~T \cdot T - 
      {\sqrt \frac{8}{9\alpha'}} ~T \cdot (T \circ T) \Bigr\}~.
\end{eqnarray}
It would be more natural if we write the action totally in
terms of the non-commutative product (\ref{eq:NCP}), i.e.
\begin{eqnarray}
  \label{eq:cubicaction}
  S = - \frac 1{2g_o^2} \int d^Dx \sqrt{g-\cF}~
      \Back&\Bigl\{&\Back 
      G^{\mu\nu} \circ \partial_\mu T \circ \partial_\nu T -
      \frac {1}{\alpha'} ~T \circ T - 
      {\sqrt \frac{8}{9\alpha'}} ~T\circ T\circ T \Bigr\}.
\end{eqnarray}
The kinetic term of this action generates the equation of motion
\begin{eqnarray}
  \label{eq:eomLagrang}
  \frac 12 \frac 1{\sqrt{g-\cF}}~
  \partial_\mu \Bigl( \sqrt{g-\cF}~(G^{\mu\nu} \circ \partial_\nu T +
  \partial_\nu T \circ G^{\mu\nu}) \Bigr) + \frac 1{\alpha'} T = 0,
\end{eqnarray}
which reduces to equation (\ref{eq:eomtachyon}) because
$\frac12(G^{\mu\nu} \circ \partial_\nu T + \partial_\nu T \circ G^{\mu\nu}) 
= G^{\mu\nu} \partial_\nu T + \cO(\partial^2G)$. 
This means that the question whether one has to put the non-commutative 
product into the kinetic term or not cannot be decided at 
first derivative order. 

If we impose the background field on-shell condition (\ref{eq:eom1}),
the kinetic term of (\ref{eq:eomtachyon}) reveals a remarkable feature
of the geometry on the D-brane. Equation (\ref{eq:eom1}) implies also
$\partial_\mu (\sqrt{g-\cF}G^{\mu\nu}) = \sqrt{g-\cF} M^{\rho\sigma} (-\Gamma_{\rho\sigma}{}^{\nu} - \frac 12 H_{\rho\sigma}{}^{\nu})$,
and we are able to rewrite (\ref{eq:eomtachyon}) as
\begin{eqnarray}
  \label{eq:eomtachyon2}
  M^{\mu\nu}\nabla_\mu \nabla_\nu T - (-2\pi) T = 0,
\end{eqnarray}
where we have introduced the connection $\nabla$ that is compatible
with the bulk metric and has torsion $H$
\begin{eqnarray}
  \label{eq:connection}
  \nabla_\mu \xi_\nu = \partial_\mu \xi_\nu
                      - \Gamma_{\mu\nu}{}^\rho \xi_\rho
                      - \frac 12 H_{\mu\nu}{}^\rho \xi_\rho .
\end{eqnarray}
This is exactly the connection that appears in closed string theory
and it is independent of the gauge field $A$.

Finally, we make a remark on higher order interactions in the
tachyonic potential. Since each term in the potential is a power of
the field $T$, it is a very symmetric expression and one may ask if
more brackets than the outermost can be omitted and, if so, how many. 
For ``$T^{\circ n}$'' with $n = 4,5$ it is easy to show that all
brackets can be left out, i.e. 
$\int_x T^{\circ 4}=\int_x T\circ T\circ T\circ T$ and 
$\int_x T^{\circ 5}=\int_x T\circ T\circ T\circ T\circ T$. 
What happens if we vary these expressions? To
this end one has to select an arbitrary choice for the brackets in
$T^{\circ 4}$ and $T^{\circ 5}$. Independently of that choice the 
variation gives the sum over all bracket arrangements for three and four
functions,
\begin{eqnarray}
  \label{eq:vary}
  && \delta \int _x T^{\circ 4} = 
     \int _x \delta T \cdot \Bigl(T\circ (T\circ T)) + 
                            ((T\circ T)\circ T)\Bigr)
\\
  && \delta \int _x T^{\circ 5} = 
     \int _x \delta T \cdot \Bigl(T \circ (T \circ (T \circ T)) +
                                  T \circ ((T \circ T) \circ T) +
                                  (T \circ T) \circ (T \circ T) +
\nonumber\\ 
  && \hspace{2.8truecm}           +~(T \circ (T \circ T)) \circ T +
                                  ((T \circ T) \circ T) \circ T \Bigr).
\end{eqnarray}
For higher powers the behaviour is different. For instance, in case of
$n=6$ the following four expressions have different variations: 
\begin{eqnarray}
  \label{eq:power6}
  &&\int _x (T \circ (T \circ T)) \circ ((T \circ T) \circ T) ,\\
  &&\int _x (T \circ (T \circ T)) \circ (T \circ (T \circ T)) ,\nonumber\\
  &&\int _x ((T \circ T) \circ T) \circ ((T \circ T) \circ T) ,\nonumber\\
  &&\int _x ((T \circ T) \circ (T \circ T)) \circ (T \circ T) .\nonumber
\end{eqnarray}
Any other bracket arrangements can be converted to one of these
possibilities by means of equations (\ref{eq:ordprod}) and
(\ref{eq:assocprod}). 
The variations of (\ref{eq:power6}) give a sum of
6, 3, 3, and 2 different terms, respectively.\footnote{%
  The total number of terms is always 14, 
  which is the number of bracket arrangements for five functions.
}
Therefore, for higher powers than five we have distinct expressions, 
where either one could arise in the tachyon potential, possibly with 
different weights.

\section{Discussion}
\label{sec:disc}

In view of the quantity $M_{\mu\nu}$ and its inverse, the two metrics,
$g$ and $G$, seem to be on equal footing on the D-brane. For instance,
the integration measure can be written as $\root 4 \of {g} \root 4 \of {G}$. 
However, in the effective field theory on the brane they play
different roles. $G$ appears in the kinetic term of the 
action and, therefore, one expects that it is the preferred metric on
the brane. However, as we have seen in section (\ref{sec:tachyon}),
the natural connection on the D-brane is compatible with
$g$ and has torsion $H=dB$. This has an interesting consequence.
The connection and the parallel transport is independent
of the open string gauge field $A$ and depends only on bulk quantities.
So we have gained some insight into the differential structure on
a D-brane and it would be even more interesting to see how 
this extends to second derivative order. The correlation functions of
open string photon vertices instead of tachyon vertices would be another
source of information. They should give rise
to a gauge theory in a general non-commutative background.

We have seen that the form of the terms in the tachyonic potential is
determined for powers lower than six. For higher powers 
of the tachyon field one has to calculate the
corresponding correlators in order to decide with which relative
weight the distinct subsets (e.g. (\ref{eq:power6})) appear. 
Of course, a sum over all bracket arrangements would be a natural choice.
But in fact, to find out whether this
guess is correct, one needs a better understanding of the underlying
$A_\infty$-structure (cf. \cite{stasheff,Cornalba:2002sm}).

Since our results are consequences of restricting off-shell correlators 
to the known structure of conformal field theory it would be
interesting to compare our results with other off-shell approaches, 
e.g. background independent open string field theory
\cite{Witten:1992qy,Witten:1993cr,
Shatashvili:1993kk,Shatashvili:1993ps}.\footnote{%
  BSFT requires a fixed conformal background in the bulk, 
  but allows for arbitrary boundary interactions. Our approach does 
  not make reference to such a background and is closer to the spirit of the 
  sigma model approach to string theory~\cite{Tseytlin:1989rr}.
} 
Clearly we expect that the tachyonic on-shell condition obtained in this 
paper is equivalent to the consistency condition for a string propagating 
in nontrivial background fields, i.e. the Weyl invariance condition of the 
underlying 2-d non-linear sigma model. Furthermore, the relation to previous 
work on the appearance of a non-commutative tachyon action, mostly considered 
in the limit of a strong magnetic field~\cite{Cornalba:2001ad,Okuyama:2001ch}, 
needs clarification. The exact non-commutative tachyon potential 
should be given by the well known expression obtained from BSFT but with 
ordinary products replaced with the generalized non-commutative product found 
in~\cite{Herbst:2001ai}.

\emph{Acknowledgments.} This work was supported in part by the Austrian 
Research Funds FWF under grants Nr. P14639-TPH and P15553,
and by the city of Vienna under grant Nr. H-85/2001.


\newpage

\thebibliography{99}\addtolength{\itemsep}{-7pt}

\vspace{-4truemm}

\bibitem{Schomerus:1999ug}
V.~Schomerus,
``D-branes and deformation quantization,''
JHEP {\bf 9906} (1999) 030
[hep-th/9903205].

\bibitem{Seiberg:1999vs}
N.~Seiberg and E.~Witten,
``String theory and noncommutative geometry,''
JHEP {\bf 9909} (1999) 032
[hep-th/9908142].

\bibitem{Ardalan:1999ce}
F.~Ardalan, H.~Arfaei and M.~M.~Sheikh-Jabbari,
``Noncommutative geometry from strings and branes,''
JHEP {\bf 9902} (1999) 016
[hep-th/9810072].

\bibitem{Ardalan:2000av}
F.~Ardalan, H.~Arfaei and M.~M.~Sheikh-Jabbari,
``Dirac quantization of open strings and noncommutativity in branes,''
Nucl.\ Phys.\ B {\bf 576} (2000) 578
[hep-th/9906161].

\bibitem{Chu:1999qz}
C.~Chu and P.~Ho,
``Noncommutative open string and D-brane,''
Nucl.\ Phys.\ B {\bf 550} (1999) 151
[hep-th/9812219].

\bibitem{Kontsevich:1997vb}
M.~Kontsevich,
``Deformation quantization of Poisson manifolds, I,''
q-alg/9709040.

\bibitem{Cattaneo:2000fm}
A.~S.~Cattaneo and G.~Felder,
``A path integral approach to the Kontsevich quantization formula,''
Commun.\ Math.\ Phys.\  {\bf 212} (2000) 591
[math.qa/9902090].

\bibitem{Herbst:2001ai}
M.~Herbst, A.~Kling and M.~Kreuzer,
``Star products from open strings in curved backgrounds,''
JHEP {\bf 0109} (2001) 014
[arXiv:hep-th/0106159].

\bibitem{Cornalba:2002sm}
L.~Cornalba and R.~Schiappa,
``Nonassociative star product deformations for D-brane worldvolumes in curved 
backgrounds,''
Commun.\ Math.\ Phys.\  {\bf 225} (2002) 33
[arXiv:hep-th/0101219].

\bibitem{Ho:2000fv}
P.~Ho and Y.~Yeh,
``Noncommutative D-brane in non-constant NS-NS B field background,''
Phys.\ Rev.\ Lett.\  {\bf 85} (2000) 5523
[hep-th/0005159].

\bibitem{Chu:2000gi}
C.~Chu and P.~Ho,
``Constrained quantization of open string in background B field and  
noncommutative D-brane,''
Nucl.\ Phys.\ B {\bf 568} (2000) 447
[hep-th/9906192].

\bibitem{Ho:2001qk}
P.~Ho,
``Making non-associative algebra associative,''
hep-th/0103024.

\bibitem{Ho:2001fi}
P.~Ho and S.~Miao,
``Noncommutative differential calculus for D-brane in non-constant B  
field background,''
hep-th/0105191.

\bibitem{stasheff}
J. D. Stasheff, 
``On the homotopy associativity of H-spaces, I. \& II.,'' 
Trans. Amer. Math. Soc. {\bf 108} (1963) 275 \& 293.

\bibitem{Polchinski:1998rq}
J.~Polchinski,
``String theory. Vol. 1: An introduction to the bosonic string,''
Cambridge University Press (1998).

\bibitem{Witten:1992qy}
E.~Witten,
``On background independent open string field theory,''
Phys.\ Rev.\ D {\bf 46} (1992) 5467
[arXiv:hep-th/9208027].

\bibitem{Witten:1993cr}
E.~Witten,
``Some computations in background independent off-shell string theory,''
Phys.\ Rev.\ D {\bf 47} (1993) 3405
[arXiv:hep-th/9210065].

\bibitem{Shatashvili:1993kk}
S.~L.~Shatashvili,
``Comment on the background independent open string theory,''
Phys.\ Lett.\ B {\bf 311} (1993) 83
[arXiv:hep-th/9303143].

\bibitem{Shatashvili:1993ps}
S.~L.~Shatashvili,
``On the problems with background independence in string theory,''
arXiv:hep-th/9311177.

\bibitem{Cornalba:2001ad}
L.~Cornalba,
``Tachyon condensation in large magnetic fields with background  
independent string field theory,''
Phys.\ Lett.\ B {\bf 504} (2001) 55
[arXiv:hep-th/0010021].

\bibitem{Okuyama:2001ch}
K.~Okuyama,
``Noncommutative tachyon from background independent open string field 
theory,''
Phys.\ Lett.\ B {\bf 499} (2001) 167
[arXiv:hep-th/0010028].

\bibitem{Tseytlin:1989rr}
A.~A.~Tseytlin,
``Sigma Model Approach To String Theory,''
Int.\ J.\ Mod.\ Phys.\ A {\bf 4} (1989) 1257.
\end{document}